\documentclass[useAMS,usenatbib]{mn2e}
\usepackage{latexsym}
\usepackage{makeidx}
\usepackage{cite}
\usepackage{amsmath}
\usepackage{color}
\usepackage{upgreek}
\usepackage{morefloats}
\usepackage{lineno}
\usepackage{setspace}
\usepackage{graphicx}

\title[Constraining the atmosphere of GJ 1214b using an optimal estimation technique]{Constraining the atmosphere of GJ 1214b using an optimal estimation technique}
\author[J.~K. Barstow et al.]{J.~K. Barstow$^{1,2}$\thanks{E-mail:
j.barstow1@physics.ox.ac.uk (JKB)}, S. Aigrain$^{1}$,
P.~G.~J. Irwin$^{2}$, L.~N. Fletcher$^{2}$, J.-M. Lee$^{2,3}$\\
$^{1}$Astrophysics, Denys Wilkinson Building, University of Oxford,
Oxford OX1 3RH, UK\\
$^{2}$Atmospheric, Oceanic and Planetary Physics, Clarendon
Laboratory, University of Oxford, Oxford OX1 3PU, UK\\
$^{3}$Institute for Theoretical Physics, University of Z\"urich,
CH-8057 Z\"urich, Switzerland}
\begin{document}

\date{Submitted April 2013}

\pagerange{\pageref{firstpage}--\pageref{lastpage}} \pubyear{2013}

\maketitle

\label{firstpage}

\begin{abstract}
We explore cloudy, extended H$_2$-He atmosphere scenarios for the warm super-Earth GJ
1214b using an optimal estimation retrieval technique. This planet,
orbiting an M4.5 star
only 13 pc from the Earth, is of particular interest because
it lies between the Earth and Neptune in size and may be a member of a new class of
planet that is neither terrestrial nor gas giant. Its relatively
flat transmission spectrum has so far made atmospheric
characterisation difficult. The NEMESIS
algorithm \citep{irwin08} is used to explore the degenerate
model parameter space for a cloudy, H$_2$-He-dominated atmosphere
scenario. Optimal estimation is a data-led approach that allows
solutions beyond the range permitted by \textit{ab initio}
equilibrium model
atmosphere calculations, and as such prevents restriction from prior
expectations. We show that optimal estimation retrieval is a powerful tool for
this kind of study, and present an exploration of the degenerate
atmospheric scenarios for GJ 1214b. Whilst we find a family of solutions that provide a very
good fit to the data, the
quality and coverage of these data are insufficient for us to more
precisely determine the abundances of cloud and trace gases given an H$_2$-He
atmosphere, and we also cannot rule out the possibility of a high
molecular weight atmosphere. Future ground- and
space-based observations will provide the opportunity to confirm or
rule out an extended H$_2$-He atmosphere, but more precise constraints
will be limited by intrinsic degeneracies in the retrieval problem,
such as variations in cloud top pressure and temperature.

\end{abstract}


\maketitle

\section{Introduction}
In recent years, the discovery of exoplanets in
close-by systems has allowed the first attempts at characterising
their atmospheres. Planets with
a high planet:star surface area ratio are the most favourable
targets, and the super-Earth sized
planet orbiting the M4.5 star GJ 1214, only 13 pc distant from the
Earth, is one such case. This planet is of particular interest because
it lies between the Earth and Neptune in size, meaning that there is
no solar system analogue and so it may be a member of a new class of
planet that is neither terrestrial nor gas giant. 

GJ 1214b was discovered in 2009 \citep{charb09} by the MEarth project
\citep{irwinj09}, a survey designed to detect any transits occurring
within a sample of 2000 nearby M dwarf stars. Radial velocity data were
obtained subsequently with the HARPS instrument, confirming
the planetary nature of the transiting object and placing a constraint
on its mass. It was found to have a radius of 2.68$\pm$0.13$R_{\earth}$ and a mass of
6.55$\pm$0.98$M_{\earth}$; these estimates were subsequently recalculated by
\citet{harps13} but both sets of values were found to be in good
agreement within the error bars. The planet's calculated density
indicates that its composition lies somewhere between a `water-world'
and  a `mini-neptune', and its equilibrium temperature is
expected to be around 550 K assuming a low Bond albedo or 400 K with a Bond albedo of 0.75
\citep{charb09}. Because its density is compatible with a range of
bulk compositions, constraining the atmospheric composition for GJ
1214b is a
crucial step towards understanding its formation process and history.

The technique of transit spectroscopy \citep{coust97,seager00} has been used with some success
to draw inferences about the atmospheres of hot Jupiter-size planets
such as HD 189733b and HD 209458b
(e.g. \citealt{knutson08,sing08,swain09,pont12}). The absorbing
species in a planet's
atmosphere can be identified by observing the transit over a range
of wavelengths; the presence of an absorber is indicated by a deeper reduction in flux
at the location of absorption features. The
shape and size of these features provide information about the
atmospheric scale height, volume mixing ratio of absorbers and the
presence of cloud or aerosol species. Further information can be
obtained when the planet is eclipsed by the star; the difference
between in and out of transit fluxes at each wavelength gives the
emission spectrum of the planet's dayside, which as well as providing
information about absorbing gases can place constraints on the
temperature structure \citep{lee12}.

This technique has also been applied to GJ 1214b, using a wide range
of ground- and space-based instruments. GJ 1214b is too cool for its
thermal emission relative to the stellar flux to be observed with
currently available telescopes \citep{mrk10}, so these
observations have been confined to measurements of transmission
through the atmosphere in primary transit. The combined efforts of
several groups have yielded a fairly continuous spectrum between 0.4 and 5 microns, making GJ 1214b one of the
best-studied exoplanets; however, the interpretation of this spectrum
has proved challenging due to a lack of significant features
\citep{bean10,bean11,berta12,demooij12}. 

The flatness of the spectrum
has led to the formulation of two competing atmospheric
models. The first, favoured by \citet{bean11}, \citet{desert11} and
\citet{berta12}, is a planet with a high molecular weight
atmosphere. All these authors produce synthetic spectra based on
an atmosphere with varying proportions of H$_2$, He and H$_2$O,
with models containing more than 50\% H$_2$O providing the best fit to
the data. An atmosphere with high molecular weight has a small
atmospheric scale height, which acts to reduce the size of absorption
features seen in transmission. The second model is a planet with a roughly solar
composition atmosphere (mostly H$_2$ and He with trace amounts of
other species) but with an opaque haze or cloud layer at high altitude that
prevents transmission through the atmosphere in between molecular
absorption bands and so masks expected features
\citep{mrk12,demooij12,howe12}. The most exhaustive range of models is
provided by \citet{howe12}. Three of their five best-fit models are
H$_2$-He atmospheres with different sizes of hydrocarbon haze particles,
based on chemicals called `tholins' that are found on Saturn's moon
Titan, and the others are N$_2$-H$_2$O-dominated atmospheres that
produce flat spectra because of their high moelcular weight. Similar
results are presented by \citet{morley13}; they use an \textit{ab
  initio} modelling technique to produce photochemical hydrocarbon
hazes in both solar metallicity and enhanced metallicity models, and
they find that hydrocarbon hazes can produce a good fit to the data in
an enhanced metallicity (50$\times$ solar) model without the
introduction of any other scattering cloud species. An extinction-only
approximation was adopted for the scattering behaviour of the
particles. 

The attempts by the above authors to fit the spectrum of GJ 1214b have been
undertaken by comparing a range of synthetic spectra generated from chemically and thermally likely model atmospheres to the data, and calculating a
goodness of fit parameter for each. This `bottom-up' approach is
useful because of its reliance on known physics of planetary
atmospheres, but exoplanets like GJ
1214b exist in regimes outside our current experience, so the
assumptions that go into these atmospheric models may be
incorrect in this case. Even for solar system planets, observations can
be very far from scenarios predicted by thermal and photochemical
equilibrium models. For example, \citet{atreya05} state that extremely patchy
ammonia cloud coverage on Jupiter seen in data from the Galileo
satellite is at odds with the global coverage expected from cloud
physics models; they suggest that spectral signatures of expected
cloud components may be masked by other hazes. Therefore, it is worth considering solutions from a
data-led approach -- optimal estimation. This approach, described in
more detail in Section~\ref{retrieval}, makes fewer assumptions about
the nature of the atmosphere and so can find atmospheric solutions
that fit the data but might be excluded by too-stringent assumptions
in \textit{ab initio} approaches.

This paper makes use of the available spectroscopic data to
investigate what constraint, if any, it is possible to place on GJ
1214b's atmosphere using the optimal estimation technique previously
presented by \citet{lee12}. The sources and treatment of the data are
discussed in Section~\ref{data}; the method and model are briefly described in
Section~\ref{retrieval}; the atmospheric models that produce
the best fit to the spectrum are presented in Section~\ref{results},
and in Section~\ref{discussion} we investigate the degeneracy and
reliability of this solution, and the improvement we expect from
future observations.

\section{Spectroscopic data}
\label{data}
Analysis of GJ 1214b is complicated by the fact that the data come
from several different sources and were obtained at different
times, as has been found by e.g. \citet{bean11}. The sources of data, wavelength ranges, instruments used and
any modifications to the errors are
listed in Table~\ref{datatable}. 

\begin{table*}
\begin{minipage}{126mm}
\centering
\begin{tabular}[c]{|l|l|l|l|}
\hline
Author & Instrument & Wavelength range & Error increase\\
\hline
\citet{bean10} & VLT/FORS Red & 0.78---1.0 $\upmu$m & 1.5$\times$\\
\citet{bean11} & VLT/FORS Blue & 0.61---0.85 $\upmu$m & 1.5$\times$\\
{}&Magellan/MMIRS & J, H, K & None\\
{}& VLT/HAWK-I & K & None\\
\citet{desert11} & Spitzer/IRAC & 3.6, 4.5 $\upmu$m & None\\
\citet{croll11} & Canada-French-Hawaii Telescope/WIRCam & J, Ks &
3$\times$ (Ks)\\
\citet{demooij12} & MPI-ESO 2.2 m/GROND & g, r, i, z & None\\
{}& Isaac Newton Telescope/WFC & r, l & 1.5$\times$ (r) \\
{}& Nordic Optical Telescope/NOTCam & Ks & 1.5$\times$\\
{}& William Herschel Telescope/LIRIS & Kc & None\\
\citet{berta12} & HST/WFC3 & 1.3---1.6 $\upmu$m & None\\
\citet{narita12} & IRSF/SIRIUS & J, H, K & 1.5 $\times$ (J) \\
\hline
\end{tabular}
\caption{Sources of spectroscopic data for GJ 1214b.\label{datatable}}
\end{minipage}
\end{table*}

\begin{figure*}
\centering
\includegraphics[width=0.8\textwidth]{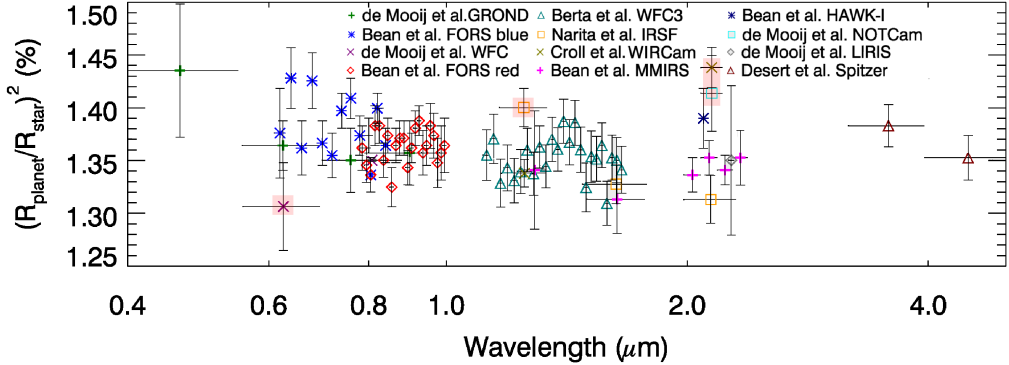}
\caption{The spectroscopic data available for GJ 1214b from different
  instruments. It can be seen that the FORS blue and red data appear
  to disagree with each other in the region of overlap, and the WFC
  r, IRSF J, and WIRCam and NOTCam Ks points are outliers (shaded in
  red). The error bars shown here are as given in the literature.\label{data_plot}}
\end{figure*}

Upon plotting these data, it becomes immediately apparent that
measurements made in the same part of the spectrum by different
instruments are not compatible within the error bars (Figure~\ref{data_plot}). This is clearly
a problem when trying to find an atmospheric model that produces a
good fit to the spectrum; for example, \citet{bean11} shift all the
VLT/FORS blue data points down by a constant factor to make them more
comparable to the VLT/FORS red points, whereas
\citet{berta12} do not do this in their analysis. In this work, rather
than artificially shift any of the data, we have chosen to increase
the error bars on some points from the estimates provided in the
original papers. This ensures that measurements made in the same
wavelength ranges with different instruments are compatible within the
error bars. Based on the decision of \citet{bean11} to shift the FORS blue data to
match FORS red, we have increased the error bars of all the FORS data
by a factor 1.5; this makes the overlapping points more comparable. We have also increased the errors on some of the broader
band measurements: the INT/WFC point
measured by \citet{demooij12} in the r-band, which has a poor
lightcurve fit, to ensure compatibility
with the ESO 2.2m telescope GROND result presented in the same paper and the
\citet{bean11} FORS blue results; the J-band point
measured by \citet{narita12} to bring it more in
line with the \citet{berta12} HST/WFC3 data; the Nordic Optical
Telescope NOTCam and Canada-French-Hawaii Telescope/WIRCam Ks
points, which are much higher than the Magellan/MMIRS, VLT/HAWK-I and IRSF/SIRIUS
values at the same wavelength. All these errors are also increased by
a factor 1.5, except that of the \citet{croll11} WIRCam Ks point
which is increased by a factor 3 since it seems to be an extreme
outlier when compared with other data. Increasing the error on data
that appear to be less reliable means that these points are given less
weight by the retrieval algorithm, to ensure that the solution is
driven by more robust observations. 

Increasing the error bars is of course not the only way of accounting
for the intrinsic disagreements within the GJ 1214b dataset (e.g. the
method of \citealt{bean11} as mentioned above) and we
explore other possibilities in Section~\ref{datausage}. This is a good
test of the robustness of our result in relation to the uncertainties
inherent in the combination of sparse, temporally separated
spectroscopic and photometric measurements. If the
  only systematics present are offsets between measurements from
  different instruments, then the method adopted here may risk
  lowering the significance of certain features within a given spectroscopic dataset unneccessarily;
  however, since we do not know the form of any potential systematics
  at this stage, we feel that this approach is reasonable in light of
  the tests described in Section~\ref{datausage}.

GJ 1214b was originally classified as an inactive M dwarf by
\citet{hawley96}, but subsequently \citet{kundurthy11} have seen hints
of spot crossing events and \citet{murgas12} see photometric
dispersion in H$\upalpha$ transit lightcurves, both of which provide
tentative evidence for stellar activity. However, \citet{berta11} find that significant
effects on transmission spectra from stellar activity are unlikely
given the current level of precision on existing data; it is likely to
become important for future observations with higher signal to
noise. Whilst \citet{demooij12} consider the effect of unocculted
star spots to be potentially important, they conclude that this is not
a suitable explanation for the observed increase in planetary radius
in the $g$- and $K_s$-bands. We cannot rule out the possibility that periods of high
activity may be responsible for the discrepancies between different
observations, but without detailed information about activity levels increasing the error
bars on these points is the most appropriate way to account for the
effect. Stellar activity monitoring and starspot correction of the kind performed for HD
189733b \citep{pont12} will have a crucial role in future observation
of GJ 1214b. 

\section{The retrieval method}
\label{retrieval}
\subsection{NEMESIS}
\label{nemesis}
NEMESIS, the Non-linear optimal Estimator for MultivariateE spectral
analySIS \citep{irwin08}, was originally developed to analyse remote sensing data from
solar system planets collected by orbiters such as Cassini and Venus
Express as well as ground-based telescope facilities. More recently,
it has been modified to allow the simulation of primary transit and
secondary eclipse spectra for extrasolar planets \citep{lee12}. Its
track record in solar system studies \citep{tsang10,irwin11,barstow12,cottini12}, versatility and efficient approach
to radiative transfer calculation make it a useful and reliable tool
for exoplanet science. NEMESIS uses the
correlated-k approximation \citep{goodyyung,lacis91},
which allows absorption coefficients over a spectral interval to be
pre-tabulated, to rapidly calculate synthetic spectra based on model
atmosphere parameters. Whilst it relies on the assumption that absorption line
strengths are well-correlated between model atmospheric layers,
i.e. lines that are strongest in the lowest atmospheric layer are also
strong in the layer above, the approach significantly reduces
computation time over the line-by-line method. Comparisons with
line-by-line calculations show that the correlated-k approximation is
sufficiently accurate for planetary atmospheric modelling
\citep{irwin08}. 

The fast forward model calculation is coupled with an optimal
estimation scheme based on the approach of \citet{rodg00}. The user
provides NEMESIS with an initial atmospheric state and an associated
error on each of the parameters to be varied - the \textit{a priori}
solution - which acts to prevent overfitting and stops retrieval
solutions from becoming unphysical. To ensure that a global solution
is found, the retrieval should be performed for a range of different
\textit{a priori} values; if the solution is global and
non-degenerate, it should be the same regardless of the initial
atmospheric state. NEMESIS then calculates a
synthetic spectrum from this initial state, and the
difference between the measured spectrum and this synthetic. It also
calculates the derivative of the spectrum with respect to
each of the variable parameters in the model, which allows an
efficient exploration of the parameter space. The best-fit solution is
found by iterating from the \textit{a priori} solution until the cost function, which represents the
difference between the measured and synthetic spectra together with
the deviation from the \textit{a priori} solution, is minimised. For
further details about the structure of NEMESIS and its use for
retrievals of extrasolar planet atmospheres, see \citet{irwin08} and
\citet{lee12}. 

There are 6 independent variables in our retrieval of GJ 1214b. We
include two populations of cloud particles and vary the total optical
depth of each; we also retrieve altitude-invariant volume mixing
ratios (VMRs) for H$_2$O, CO$_2$ and CH$_4$, and the radius of the
planet at the 10-bar level. This last parameter is necessary because the radius quoted in the
literature is derived from white light transit observations and is
simply the radius at which the atmosphere becomes opaque to white
light, which is dependent on the atmospheric properties. When the
VMRs of the active gases change in the retrieval, the VMRs of the H$_2$ and He that
make up the rest of the atmosphere are scaled (with a fixed ratio of 0.85:0.13) to ensure the sum
of the VMRs is unity. 

To investigate the sensitivity of the retrieval to the chosen \textit{a
  priori} solution, we follow a `bracketed retrieval' procedure. For
each of the model parameters except the radius, which effectively just
scales the final result, we use 21 different starting points spanning
4 orders of magnitude and perform the retrieval for each of these 21
starting points. This means we run the retrieval a total of 105
times, varying the \textit{a priori} for only one parameter at a time. See Section~\ref{atmosmodel} for further
details of the \textit{a priori} choice. A reduced $\chi^2$ parameter\footnotemark is calculated in each case, with a
good fit being where the reduced $\chi^2$ is close to 1. This process
tests the sensitivity to the \textit{a priori} and the reliability of
the best-fit solution. 

\footnotetext{The reduced $\chi^2$ is the $\chi^2$ goodness-of-fit
  parameter divided by the number of degrees of freedom (number of
  spectral points - number of model variables - 1).} 

\subsection{Atmospheric model}
\label{atmosmodel}
The advantage of the optimal estimation approach is that, if there is
sufficient information available in a spectrum to constrain
atmospheric properties, the \textit{a priori} atmospheric state should
not affect the result \citep{irwin08}. The \textit{a priori} atmospheric state we
assume here is based on the best-fit models of
\citet{howe12}. \citet{howe12} demonstrate that it is possible to
produce a fit of comparable quality to a high-molecular-weight model
atmosphere by adding tholin haze particles to a H$_2$-He model
atmosphere, but their models do not successfully simultaneously fit both the slight slope in
the blue part of the spectrum (0.5---0.8 $\upmu$m, see
Figure~\ref{data_plot}) and the flatter infrared region (longwards of 0.8
$\upmu$m). 

\subsubsection{Clouds}
\citet{howe12} use
models with different sizes of tholin haze particle, but they only include one
size in each separate model. Cloud particle
extinction is maximised in the Mie scattering regime, where the particle size is
comparable to the wavelength of light, so in order to fit the apparent
increase in extinction towards the blue end of the spectrum and the
fairly constant extinction throughout the infrared a range of particle
sizes is required. We therefore adopt a 2-mode cloud model:
a very narrow size distribution of 0.1 $\upmu$m-sized particulate haze
(`cloud 1')  and a broader log-normal distribution of larger particles
with a modal radius of 1 $\upmu$m and a width of 0.25 (`cloud 2'). The narrowness of the 0.1
$\upmu$m haze size distribution means that the extinction efficiency
decreases throughout the visible, producing the blue---red downward
slope in the transmission spectrum, whilst the broader size
distribution in cloud 2 allows absorption over a broader range of
wavelengths in the infrared as the extinction efficiency does not
decrease as rapidly with wavelength. We use tholin refractive index
data from \citet{khare84}. This kind of multi-modal cloud model has
been used to successfully model the Venusian sulphuric acid haze and cloud (e.g. \citealt{crisp86a,pol93,grin93}).

It is of course possible that any clouds on GJ 1214b would be made of
something other than tholins; however, in this model we adopt a single
scattering approximation, which means that the main variable affecting
the amount of light scattered by the cloud is just the size of the
cloud particle and not what it is made of. \citet{dekok12} point out
that this approximation is not valid where particles are highly
forward or backward scattering, but given the large error bars on the
GJ 1214b spectrum we expect any uncertainty from scattering assumptions to
be second order. Tholins are also a logical choice of cloud
constituent where the optical depth is allowed to vary freely, because
they are made of hydrocarbons and H and C are widely available in a
solar composition atmosphere. Other likely constituents in the GJ 1214b
temperature range are ZnS and KCl \citep{morley12,morley13}, but the solar abundances
of the metals restrict the maximum cloud optical depth that can be
achieved.

We also do not allow the vertical positioning of the cloud to vary in
our nominal model. After \citet{howe12}, we place the cloud in the pressure
range 1-100 mbar; the second cloud population with larger particles
only extends up to 3 mbar altitude as we would not expect larger
particles to be supported up to the same altitude as the smaller
haze (e.g. as on Venus). The \textit{a priori} cloud number
densities are shown in Figure~\ref{tempprof}. We test the effect of varying the vertical positioning of the cloud in
Section~\ref{cloudalt}, by repeating the retrieval with a model
cloud top at three different altitudes and comparing the results. 

\subsubsection{Gases}
We adopt a bulk H$_2$-He atmospheric model, as \citet{howe12}, with
trace amounts of H$_2$O, CO$_2$ and CH$_4$. We restrict ourselves to
these few species as they are the most spectrally active in the region of
interest out of the common molecules we expect to occur. These are
also the mini-Neptune scenario constituents adopted by \citet{benneke12} in their paper
detailing observations of a GJ 1214b-like atmosphere with the James
Webb Space Telescope. We include
CH$_4$ but not CO because CH$_4$ should be far more abundant than CO in GJ 1214b due
to its temperature \citep{lodders02}. The sources of the absorption data for these
three gases are listed in Table~\ref{linedata}. We also include
H$_2$-He collision-induced absorption  as in \citet{lee12}.

\begin{table}
\centering
\begin{tabular}[c]{|c|c|}
\hline
Gas & Source\\
\hline
H$_2$O & HITEMP2010 \citep{roth10}\\
CO$_2$ &  CDSD-1000 \citep{tash03}\\
CH$_4$ & STDS \citep{weng98}\\
\hline
\end{tabular}
\caption{Sources of gas absorption line data.\label{linedata}}
\end{table}

\begin{table}
\centering
\begin{tabular}[c]{|c|c|}
\hline
Variable & Value\\
\hline
Cloud 1 & 10$^9$\\
Cloud 2 & 10$^6$\\
H2O VMR & 1000 \\
CO2 VMR & 100\\
CH4 VMR & 500\\
\hline
\end{tabular}
\caption{Initial number densities of cloud species (number per gram of
  atmosphere) and VMRS of
  spectrally active gases (ppmv). The cloud abundances translate to
  5600 0.1$\upmu$m particles per cm$^3$ and 5.6 larger particles per
  cm$^3$ at the cloud base pressure of 100 mbar, which then decreases with the
  atmospheric pressure scale height towards higher altitudes. The gas
  VMRs are assumed to be constant with altitude. \label{abundances}}
\end{table}

Due to the flatness of the spectrum, only upper limits on the VMRs of
H$_2$O, CO$_2$ and CH$_4$ are likely to be achievable, as this means
that the
signal-to-noise on the variation of the spectrum with wavelength is
small. 

\subsubsection{Temperature Structure}

\begin{figure}
\includegraphics[width=0.4\textwidth]{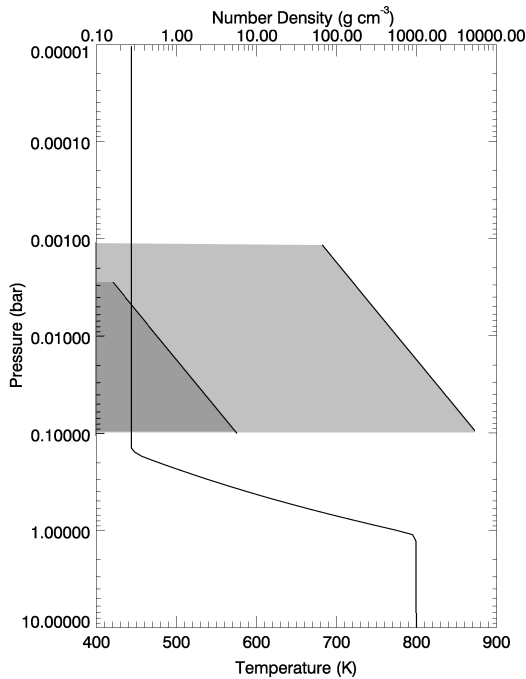}
\caption{The model temperature profile for GJ 1214b, plus \textit{a
    priori} number
  densities for the 0.1 $\upmu$m haze (light grey) and 1 $\upmu$m
  cloud (dark grey), all as a function of pressure. The model
  atmosphere extends up to 10$^{-12}$ bar, but because it is
  isothermal above 0.1 bar the upper atmosphere is not shown in this
  plot.\label{tempprof}}
\end{figure}

The temperature structure of GJ 1214b is not known, nor is there
enough information in the transmission spectrum to independently constrain it
along with all the other possible variables \citep{barstow13}. It is
therefore necessary to use an estimated temperature profile in the
atmospheric model. Whilst there isn't enough information to
independently constrain the temperature, changes in the temperature
structure do nonetheless affect the transmission spectrum because the
atmospheric scale height depends on the temperature -- 
increasing the temperature increases the scale height, and vice versa. However, increasing
the planetary radius would produce the same effect on the spectrum as increasing the temperature, because it lowers the gravitational acceleration of the
planet which also increases the scale height. \citep{barstow13}. Any
inaccuracies in our temperature estimation will therefore be
compensated for to some extent in the radius retrieval and so should not
significantly affect the result. This hypothesis is tested in
Section~\ref{tempvary}. GJ 1214b may be tidally locked as it is close to its
parent star ($\sim$0.01 AU), meaning that equilibrium temperature estimates are not
necessarily valid for the terminator regions, so this is another
reason for testing the sensitivity of the retrieval to the estimated
temperature profile.
\begin{figure*}
\centering
\includegraphics[width=0.8\textwidth]{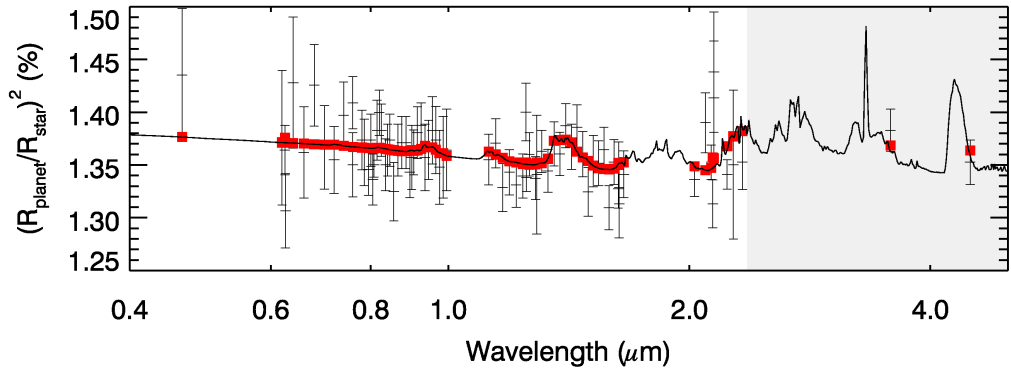}
\caption{Best-fit spectrum for GJ 1214b, based on the NEMESIS
  bracketed retrievals. The black points with error bars are the
  data, the red squares are the best-fit synthetic spectrum convolved
  with the filter functions for each data point, and the black line
  is the best-fit spectrum at R=300. The highlighted region shows
  features in a currently data-poor part of the spectrum that may
  help to distinguish between different scenarios.\label{bestfit}}
\end{figure*}

The temperature structure we adopt is arrived at using the estimation process
described in \citet{barstow13}, and is shown here in
Figure~\ref{tempprof}. The stratospheric temperature is calculated
based on an assumed equilibrium temperature of 530 K, towards the
higher end of the range indicated by \citet{charb09}, corresponding to a
Bond albedo of 0.15. We explore the effect of varying the stratospheric temperature in
Section~\ref{tempvary}. As in \citet{barstow13}, we assume the
presence of an adiabat between 1 and 0.1 bar, and we use the specific
heat capacities $c_p$ for H$_2$ and He at the stratospheric
temperature, as they do not vary greatly over the temperature range in
the model profile. The deep atmospheric temperature is calculated
using equation~\ref{radeq4}:
\begin{equation}
T_{\mathrm{trop}}=T_{\mathrm{strat}} -\frac{kT_{\mathrm{strat}}}{mc_p}\mathrm{ln}\left(\frac{p_1}{p_2}\right) \label{radeq4}
\end{equation} 
where $k$ is the Boltzmann constant, $T_{\mathrm{strat}}$ is the
stratospheric temperature and $m$ is the molecular mass of the
atmosphere. $p_1$ and $p_2$ are pressures at the top and bottom of the
adiabat.

Whilst we have explained the derivation of a temperature profile with an adiabatic troposphere, even in a
cloud-free atmosphere no signal will be observed from pressure
levels deeper than 0.1 bar in transmission \citep{barstow13}; the
details of the temperature profile below this are therefore relatively
unimportant, so it is the isothermal stratospheric temperature that we
expect to have the largest effect on the retrieval.

\section{Results}
\label{results}

Our best fit spectrum to the available GJ 1214b data is presented in Figure~\ref{bestfit}. We show it both integrated at the
resolution of the individual observation bands and at a resolving
power of 300. It can be seen that there are a series of clear
molecular absorption features
between 2 and 5 $\upmu$m, which the currently available data do not
probe in detail. If these features could be observed a much stronger
constraint on the atmosphere would be obtained. 

\begin{table}
\centering
\begin{tabular}[c]{|c|c|c|}
\hline
Variable & Value & Error \\
\hline
Cloud 1 & 1.55 & 1.00\\
Cloud 2 & 0.783  & 1.13\\
H2O VMR & 1.16 & 1.04\\
CO2 VMR & 0.876 & 1.16\\
CH4 VMR & 0.169  & 0.732\\
10-bar Radius & 15320 km & 58 km\\
\hline
\end{tabular}
\caption{The parameter values in the best-fit model; all are
  expressed as multiplying factors on the model values listed in
  Table~\ref{abundances}, except the radius which is in km. For the
  multiplying factors, the error given is the error in the logarithm.\label{bestmodel}}
\end{table}

\begin{figure*}
\centering
\includegraphics[width=1.0\textwidth]{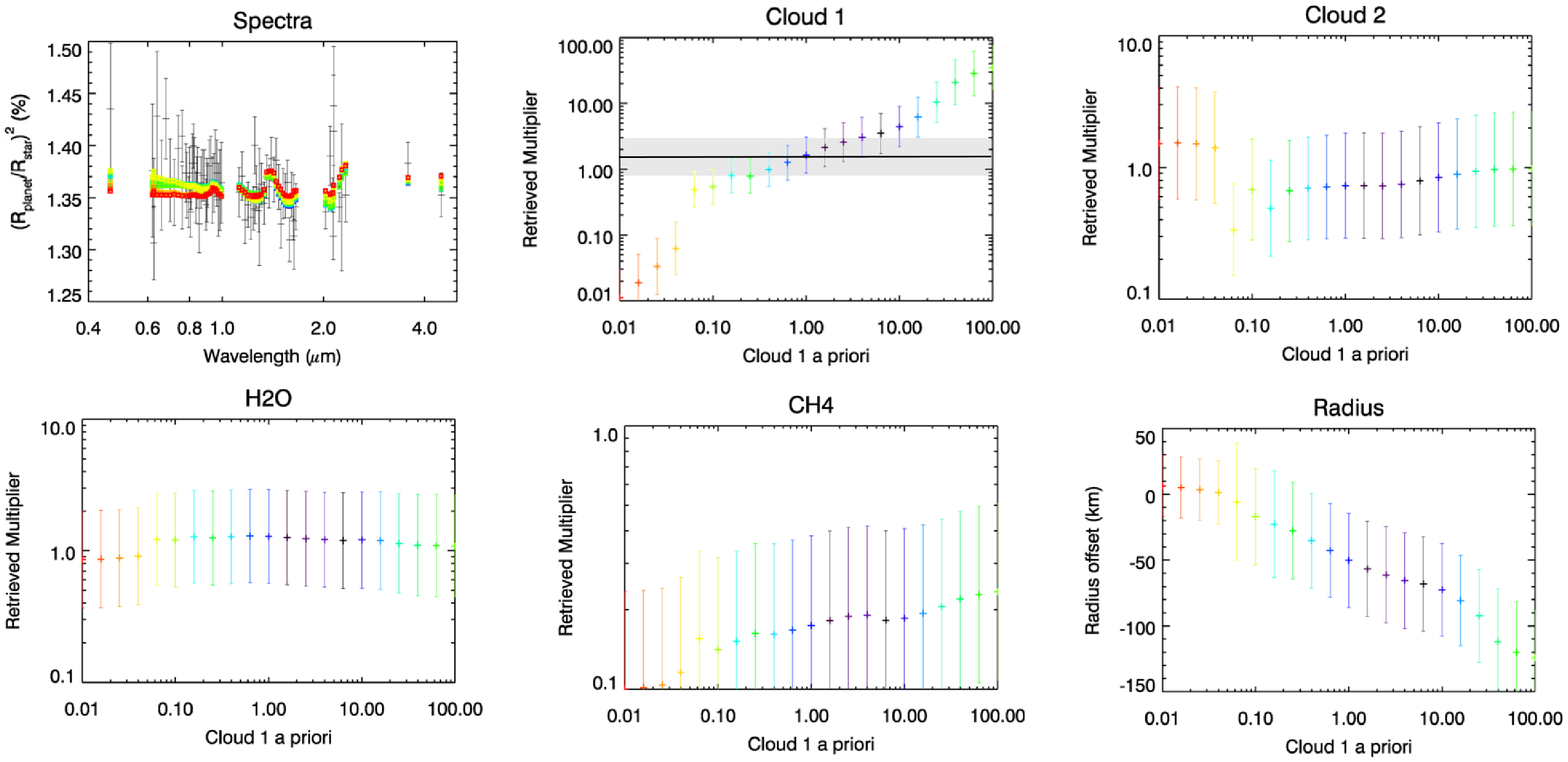}
\caption{Bracketed retrieval results for GJ 1214b, where the
  \textit{a priori} being altered is the cloud 1 number density. We
  do not plot the results for CO$_2$ as the retrieved
  values do not vary significantly for different \textit{a priori}
  scenarios. The different colours correspond to different values of
  the reduced $\chi^2$, with black lowest and red highest, but in
  this case all are $\sim$0.8--1. The best-fit value and errors for cloud
  1 are shown (black line and grey shading).\label{cloud1results}}
\end{figure*}
\begin{figure*}
\centering
\includegraphics[width=1.0\textwidth]{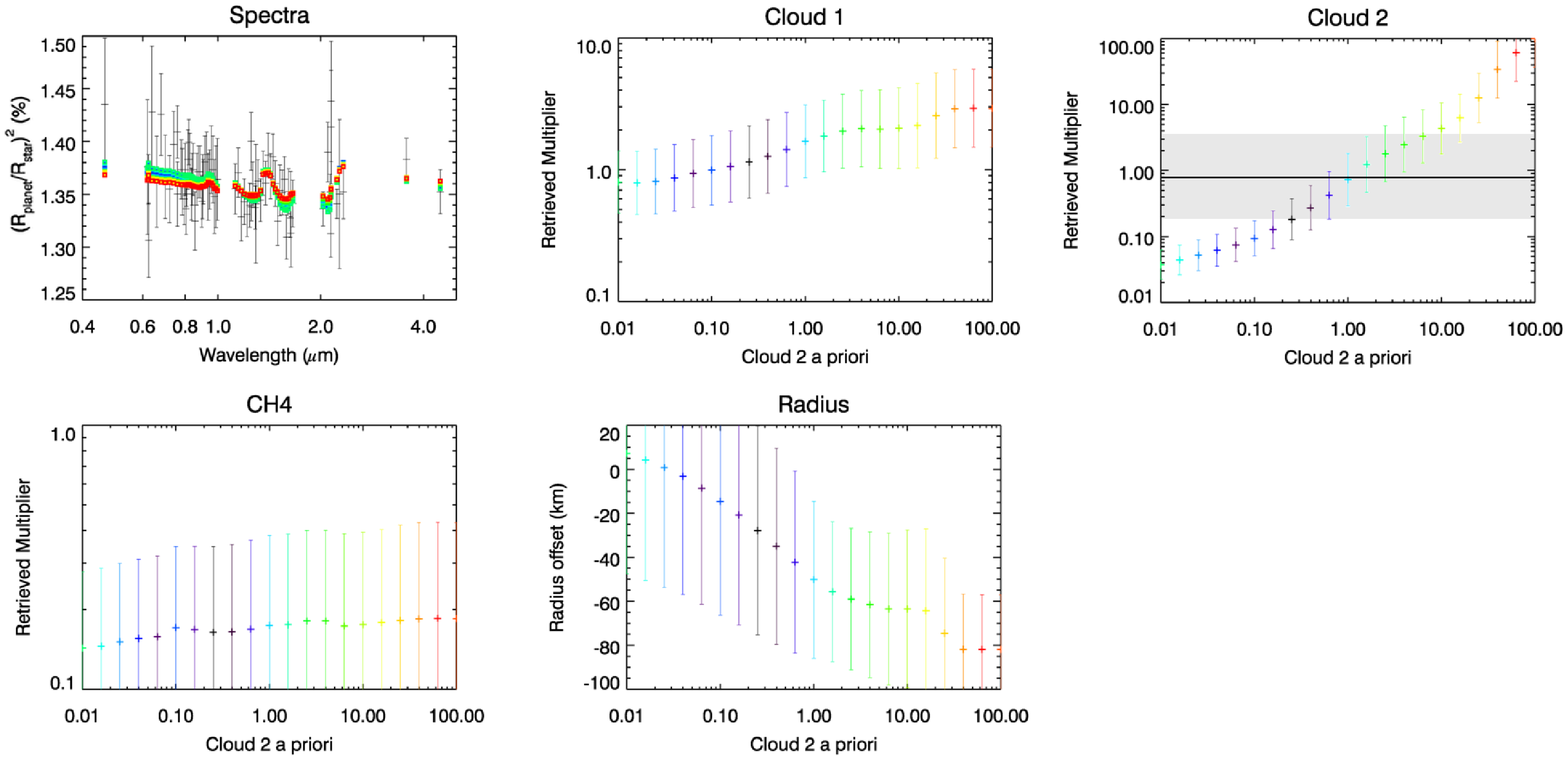}
\caption{Bracketed retrieval results for GJ 1214b, where the
  \textit{a priori} being altered is the cloud 2 number density. We
  do not plot the results for H$_2$O or CO$_2$ as the retrieved
  values do not vary significantly for different \textit{a priori}
  scenarios. The different colours correspond to different values of
  the reduced $\chi^2$, with black lowest and red highest, but in
  this case all are $\sim$0.8. The best-fit value and errors for cloud
  2 are shown (black line and grey shading). \label{cloud2results}}
\end{figure*}
\begin{figure*}
\centering
\includegraphics[width=1.0\textwidth]{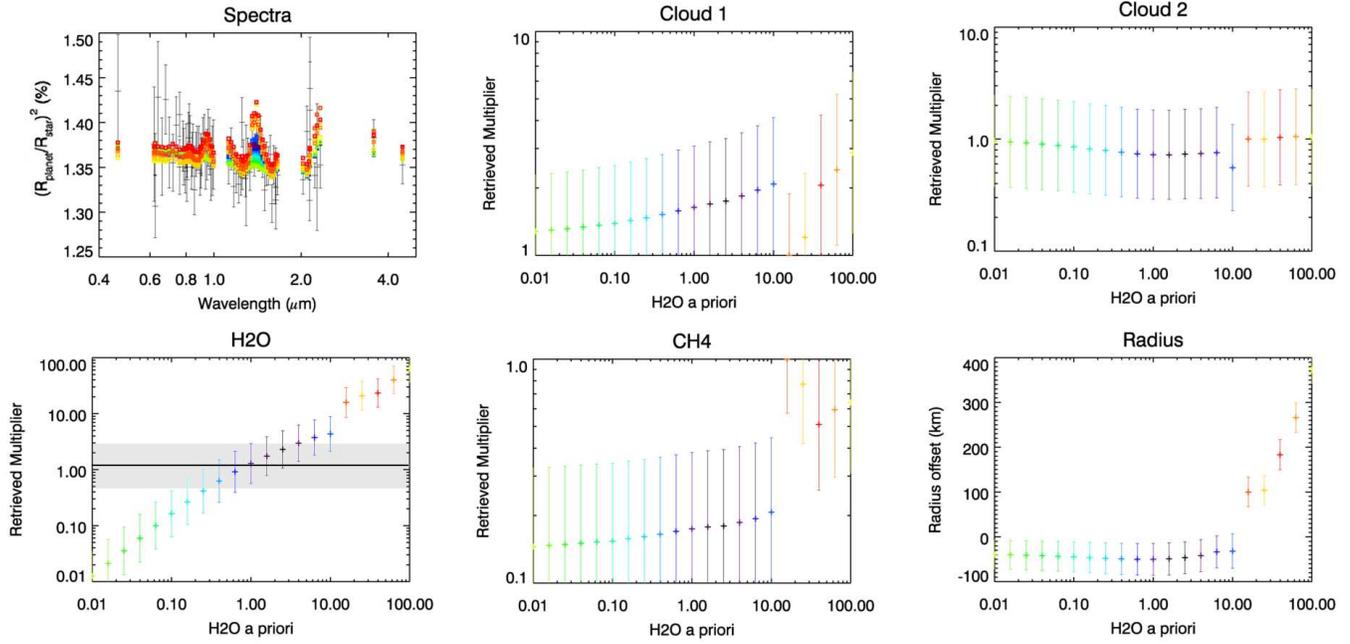}
\caption{Bracketed retrieval results for GJ 1214b, where the
  \textit{a priori} being altered is the H$_2$O VMR. We
  do not plot the results for CO$_2$ as the retrieved
  values do not vary significantly for different \textit{a priori}
  scenarios. The different colours correspond to different values of
  the reduced $\chi^2$, with black lowest and red highest, but in
  this case all except the rightmost 5 points are $\sim$0.8; those
  5 points are between 1.0 and 1.46. The best-fit value and errors
  for H$_2$O are shown (black line and grey shading).\label{h2oresults}}
\end{figure*}
\begin{figure*}
\centering
\includegraphics[width=1.0\textwidth]{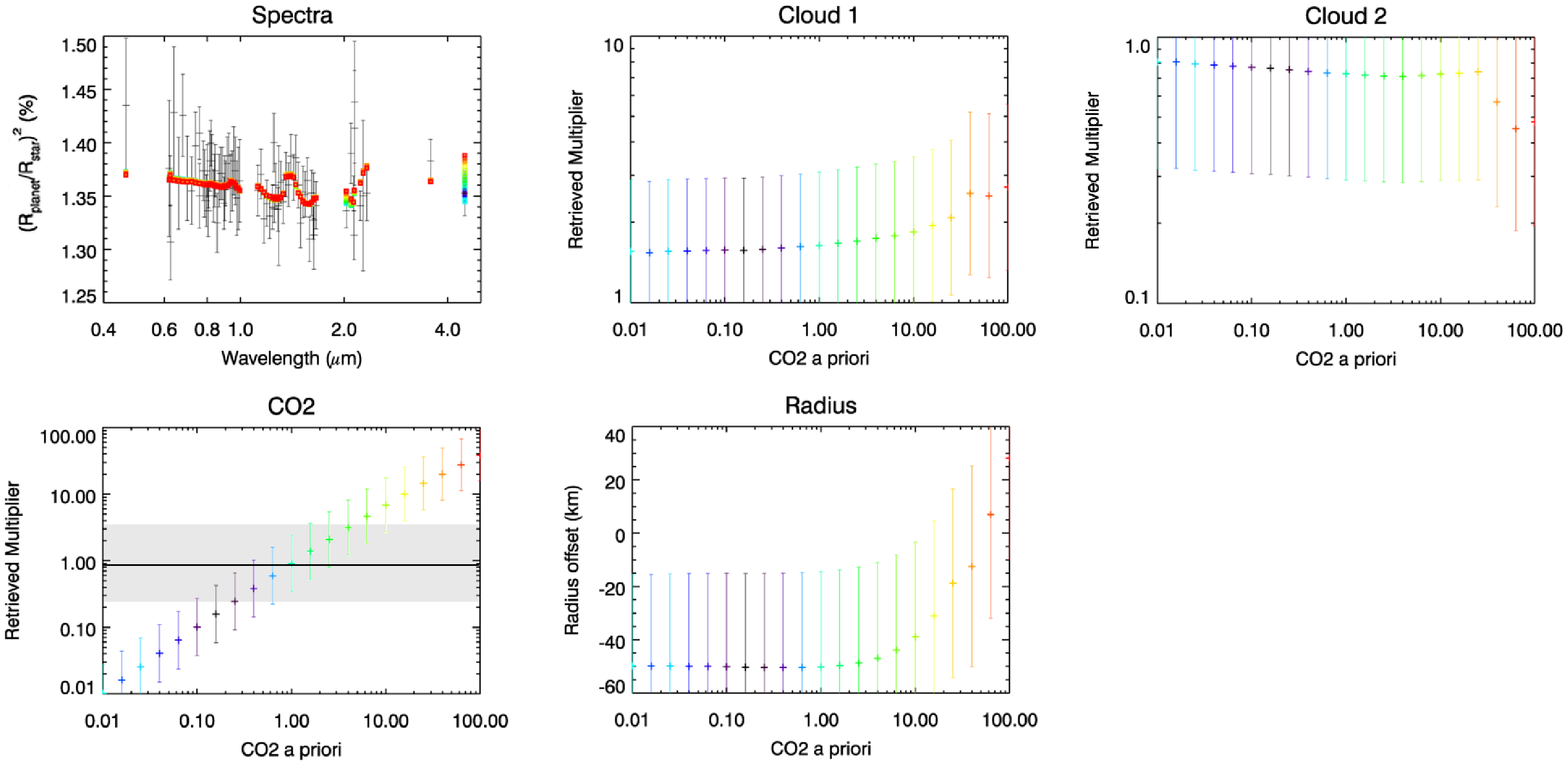}
\caption{Bracketed retrieval results for GJ 1214b, where the
  \textit{a priori} being altered is the CO$_2$ VMR. We
  do not plot the results for H$_2$O and CH$_4$ as the retrieved
  values do not vary significantly for different \textit{a priori}
  scenarios. The different colours correspond to different values of
  the reduced $\chi^2$, with black lowest and red highest, but in
  this case all are $\sim$0.8. The best-fit value and errors for CO$_2$ are shown (black line and grey shading).\label{co2results}}
\end{figure*}
\begin{figure*}
\centering
\includegraphics[width=1.0\textwidth]{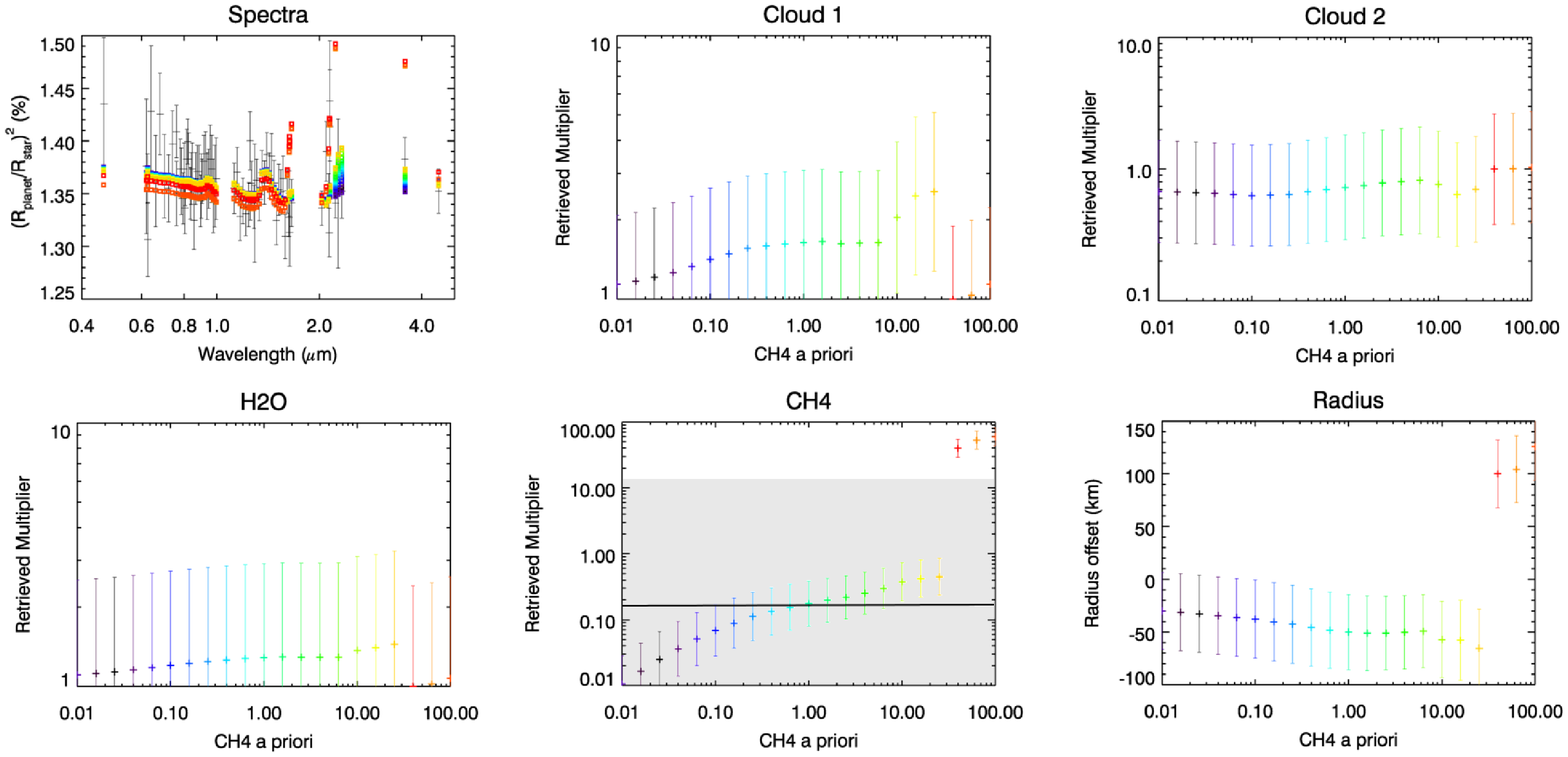}
\caption{Bracketed retrieval results for GJ 1214b, where the
  \textit{a priori} being altered is the CH$_4$ VMR. We
  do not plot the results for cloud 2 and CO$_2$ as the retrieved
  values do not vary significantly for different \textit{a priori}
  scenarios. The different colours correspond to different values of
  the reduced $\chi^2$, with black lowest and red highest, but in
  this case all are $\sim$0.8 except the rightmost three points in the
  plots, which are $\sim$4. The best-fit value and errors for CH$_4$ are shown (black line and grey shading).\label{ch4results}}
\end{figure*}

The optimal values of the model parameters and associated errors are
presented in Table~\ref{bestmodel}. It can be seen that the error bars
on these values are very large, because they represent a weighted
average over each retrieval run in the bracketed retreival test
described in Section~\ref{nemesis}. The
weighting used in this case is the calculated reduced $\chi^2$, with
larger $\chi^2$ values being given less weight since those models
produce a poorer fit to the spectrum. 

\begin{figure}
\includegraphics[width=0.5\textwidth]{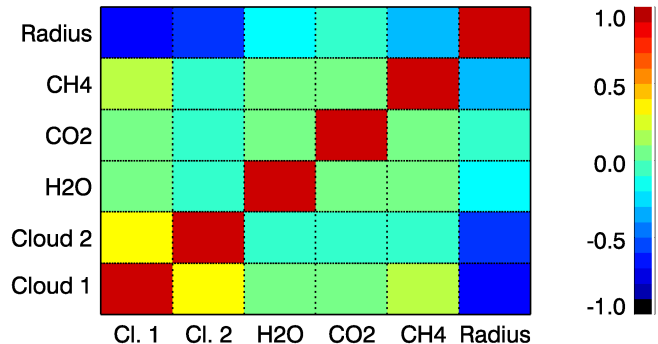}
\caption{The correlation matrix for this retrieval problem. A
  correlation of +1 indicates perfect positive correlation, -1
  perfect negative correlation. There is significant negative
  correlation between radius and all the other variables except H$_2$O
  and CO$_2$ VMR. There is significant positive correlation
  between cloud 1 number density and cloud 2 number density, and CH$_4$ VMR.\label{ctmatrix}}
\end{figure}

The full results of the bracketed retrieval test are shown in
Figures~\ref{cloud1results}---\ref{ch4results}. For each test, the variations in the retrieved
values for each variable are shown, along with the synthetic spectra produced
from each retrieval run. The colours correspond to the reduced
$\chi^2$, with the red point in each plot having the highest $\chi^2$
and the black point the lowest. In general, however, the variation in
$\chi^2$ is not very great and there are few models which give a fit
with a reduced $\chi^2$ significantly greater than 1. Several
combinations of model parameters produce an equally good fit to the
spectrum, with some values varying over several orders of magnitude,
indicating that this problem is highly degenerate; this is also clear
from Figure~\ref{ctmatrix}, which shows the correlation between the
different retrieval variables. The only variables that do not show
significant (magnitude
$>2.5$) correlation with another are the H$_2$O and CO$_2$ VMR; cloud 1 optical depth is positively
correlated with cloud 2 optical depth and CH$_4$ VMR, and
the radius is negatively correlated with cloud optical depths and
CH$_4$ VMR. The high correlations are indicative of
degeneracy between different model atmosphere scenarios; for example,
increasing the cloud optical depths increases the opacity of the
atmosphere at higher altitudes, making the planet appear bigger, but this effect can be compensated for
if the radius at 10 bar is decreased. This means that a large range of
cloudy model atmospheres are plausible,
which is similar to the result of
\citet{morley13}, who also find that several models containing
hydrocarbon hazes provide an adequate fit to the data. 

It is also clear in Figures~\ref{cloud1results}---\ref{ch4results}
that in the majority of cases the retrieval is being driven by the
chosen \textit{a priori} rather than by information in the measured
spectrum, as the retrieved parameters are clearly correlated with the
\textit{a priori}. This is particularly severe for cloud 2, H$_2$O and
CO$_2$, where the correlation is almost linear. We cannot therefore
draw any reliable conclusion from these retrieval results, although
likely values for the model parameters are indicated. We also find
that a straight line can fit the available data with a reduced
$\chi^2$ of 0.94, and therefore we cannot claim to have detected the
presence of any molecular species on GJ 1214b. GJ 1214b is
insufficiently dense to be a rocky, atmosphere-less planet, so we do
not expect that a straight line represents a realistic, physical
scenario, but this fact demonstrates the limitations of the current dataset.

\begin{table*}
\begin{minipage}{126mm}
\begin{tabular}[c]{|c|c|c|c|c|c|}
\hline
Variable & Original & Pub. errors & Spectroscopic & FORS blue down &
FORS red up \\
\hline
Cloud 1 & 1.55$\pm$1.00 & 1.38$\pm$0.93& 1.69$\pm$0.90&1.60$\pm$1.00&1.73$\pm$0.892\\
Cloud 2 &0.783$\pm$1.13 & 0.704$\pm$1.10 & 0.504$\pm$1.22&0.743$\pm$1.13 &0.489$\pm$1.22\\
H2O VMR &1.16$\pm$1.04 & 1.00$\pm$1.05 & 0.980$\pm$1.01&1.27$\pm$1.03 &0.972$\pm$1.01\\
CO2 VMR & 0.876$\pm$1.16& 0.879$\pm$1.17 & 0.838$\pm$1.12&0.856$\pm$1.14 &0.833$\pm$1.12\\
CH4 VMR &0.169$\pm$0.732  & 0.199$\pm$0.782  & 0.174$\pm$0.689&0.199$\pm$0.737&0.175$\pm$0.694 \\
10-bar Radius & 15320$\pm$58 km& 15345$\pm$58 km & 15340$\pm$53
km&15327$\pm$57 km&15342$\pm$54 km\\
\hline
\end{tabular}
\caption{The parameter values in the best-fit models for five
  different treaments of the available data.\label{datatests}}
\end{minipage}
\end{table*}

The retrieved H$_2$O volume mixing ratio shows almost a 1:1 correspondance
with the \textit{a priori} value (Figure~\ref{h2oresults}), making it impossible to
trust the retrieval for this variable. This is because of the
trade-off of two different effects governing the size of the H$_2$O
absorption features; increasing the abundance of H$_2$O increases the
absorption due to this gas (features look larger), but it also increases the molecular weight
and therefore reduces the scale height of the atmosphere (features
look smaller). A range of H$_2$O volume mixing ratios spanning 4
orders of magnitude is compatible with the observations, and it has
already been shown by several authors \citep{bean11,desert11,berta12}
that even higher abundances of H$_2$O are compatible with the data. It
is clearly not possible to place a meaningful constraint on the
abundance of H$_2$O in GJ 1214b's atmosphere with the data currently
available. This is similarly true of the CO$_2$ abundance; whilst
there is no significant degeneracy between CO$_2$ abundance and other
parameters, there is clearly a strong dependence on the \textit{a
  priori} (Figure~\ref{co2results}). We can infer from this that the
spectrum is not strongly affected by the presence of CO$_2$ in the
model atmosphere since variations in CO$_2$ abundance do not affect
the retrieval of other properties, so there is currently no evidence for its presence on GJ
1214b. 

The only variable we can place any constraint on is the CH$_4$ VMR;
where the retrieved VMR is above 20$\times$ the \textit{a priori},
equivalent to  1\%, the reduced $\chi^2$ is significantly higher
(Figure~\ref{ch4results}), so we can place a tentative upper limit of
1\% on the CH$_4$ abundance in GJ 1214b's atmosphere, within the
limitations of our model scenario. However, this is not a very
stringent constraint, and we conclude that at present there is not
enough information in the data to reliably constrain GJ 1214b's
atmosphere, and we do not claim to have detected either cloud or
molecular features in the spectrum.

\section{Discussion}
\label{discussion}
It is clear from the results presented in Section~\ref{results} that
the retrieval is heavily dependent on the \textit{a priori}
assumptions in our atmospheric model, and therefore we can arrive at
no firm conclusion about the nature of GJ 1214b's atmosphere. We have shown that it is possible to produce a good fit to the full
visible and infrared spectrum with a cloudy H$_2$-He atmosphere, but
it is important to understand the implications of the assumptions that
went into this model. We present a series of further retrieval tests
below, in which we have altered some of the non-retrieved model
parameters/data to investigate their influence on the result. Whilst
these tests do not shed any further light on GJ 1214b's atmospheric
composition at this time, in order for retrieval methods to fully exploit any future
measurements it is crucial that we understand
the sensitivities of the spectrum to the model parameters. 

\subsection{Data usage}
\label{datausage}

Since this work commenced, further observations of GJ 1214b have been
published, most notably those of \citet{fraine13}; these authors have
repeated the warm Spitzer 3.6 and 4.5 $\upmu$m measurements of \citet{desert11}, and whilst
the radii are not incompatible with those previously derived the
errors are somewhat smaller. This may result in the Spitzer points
providing further constraint on the model atmosphere, so we repeat the above analysis including
the \citet{fraine13} points instead of the \citet{desert11}
points. This produced a small (\textless10\%) difference in the values of all
best-fit model parameters except the CO$_2$ and CH$_4$ VMRs, with the
CO$_2$ multiplier reduced to 0.682$\pm1.10\times$ from 0.876$\times$ the \textit{a priori}
VMR and CH$_4$ increased to 0.237$\pm0.54\times$ from 0.169$\times$.  However, this variation
is well within the retrieval error. 

We did not test the inclusion of the \citet{fraine13} Iz (0.8---1.1
$\upmu$m) point, since
it is compatible with existing measurements in the wavelength region
and we felt it was unlikely that any information would be added.


As mentioned in Section~\ref{data}, combining data at different
wavelengths from multiple sources is often problematic in transmission
spectroscopy, because of temporal changes in stellar activity and also
different instrument systematics/processing techniques from different
observations. In our original analysis, we have attempted to account
for this by increasing the error bars on some measurements to ensure
that measurements obtained in the same wavelength region are in
agreement within their error bars. To test the impact of this, we
repeated the analysis with the published errors. We also performed the
same analysis with only the spectroscopic datasets (VLT/FORS blue and red,
HST/WFC3 and Magellen/MMIRS K-band) plus the Spitzer/IRAC measurements,
since the ground-based photometric data points are seen to be the most
discrepant (Figure~\ref{data_plot}) and are also the most difficult to
match with data at different wavelengths obtained at different times;
it is extremely challenging to absolutely calibrate the out-of-transit baseline
for atmospheric effects, adding to the uncertainty on the transit
radius for single photometric points relative to other datasets. This
test also used the published errors.

We performed two additional tests for the spectroscopic data plus
Spitzer; to check the sensitivity of the result to shifts in the
baseline radius for different datasets, we shifted the FORS blue
points down by the same amount as \citet{bean11}. We also shifted the
FORS red points up by the same amount in a separate test. The
average best-fit retrieved values and errors are shown for each of
these test cases in Table~\ref{datatests}.

It can be seen in Table~\ref{datatests} that the
variation in the average best-fit retrieved values is well within the
error bounds on those values. The behaviour of the bracketed retrieval
is also robust under the different combinations and treatments of the
data, as the same correlations between variables and dependence on the
\textit{a priori} seen in
Figures~\ref{cloud1results}---~\ref{ch4results} are reproduced in all
cases. The only difference is that the reduced $\chi^2$ is
somewhat higher for the published error case, at $\sim$1.5. Our result, namely that the existing data are non-constraining,
is therefore not dependent on the details of the datasets chosen or
the treatment of the error bars on those datasets. For future analyses of this kind, in which spectroscopic
features can be resolved with a reasonable signal-to-noise,
a more detailed approach would be necessary. An appropriate technique would be to create a
  grid of offsets between different datasets, and then to run the
  retrieval for all cases to examine the effect on the result of any unknown
  systematic errors. This could be extended to also include variable
  gradients for
  visible data, which are the most likely to be
  affected by the presence of star spots. We stress that the error
  budget for spectroscopic measurements that are combined in this way is likely to be
  dominated by systematics, so any conclusive result must involve
  rigorous testing of the kind described in order to ensure its robustness. However,
  since we cannot draw any firm conclusions from the existing GJ 1214b data, it
  is clear that further testing would serve no purpose in this case.



\subsection{Cloud altitude}
\label{cloudalt}
In the previous section, we retrieved the cloud particle abundances
for both cloud modes, but we did not allow the altitude of the cloud
to vary. In transmission geometry, the cloud top altitude is most
important because the long slant path through the cloud means that it quickly
becomes optically thick deeper in the atmosphere. We adjusted the cloud
top pressures to 0.1 mbar and 10 mbar from the \textit{a priori} of 1
mbar and repeated the retrieval analysis for these cases. A good fit
to the data can be obtained for all of these cloud top pressures, and
the best-fit parameter values for each pressure are
shown in Table~\ref{cloudtops}.

\begin{table}
\centering
\begin{tabular}[c]{|c|c|c|c|}
\hline
Variable & 10 mbar & 1 mbar & 0.1 mbar\\
\hline
Cloud 1 & 0.981$\pm$1.16 &1.55$\pm$1.00& 0.530$\pm$0.867\\
Cloud 2 & 0.301$\pm$1.59&0.783$\pm$1.13& 0.162$\pm$1.25\\
H2O VMR & 0.522$\pm$0.788& 1.16$\pm$1.04& 1.75$\pm$1.11\\
CO2 VMR & 0.619$\pm$1.00& 0.876$\pm$1.1 & 0.971$\pm$1.17\\
CH4 VMR & 0.0575$\pm$0.432& 0.169$\pm$0.732 & 0.219$\pm$0.841\\
10-bar Radius & 15515$\pm$27 km& 15320$\pm$58 km& 15224 km$\pm$97 km\\
\hline
\end{tabular}
\caption{The parameter values in the best-fit models for three
  different cloud top pressures,
  expressed as multiplying factors on the model values listed in
  Table~\ref{abundances} except for the radius which is in km.\label{cloudtops}}
\end{table}

\begin{table}
\centering
\begin{tabular}[c]{|c|c|c|c|}
\hline
Variable & -50 K & +0 K & +50 K \\
\hline
Cloud 1 & 2.06$\pm$0.97 & 1.55$\pm$1.00& 1.11$\pm$0.985\\
Cloud 2 & 0.813$\pm$1.17 & 0.783$\pm$1.13 & 0.591$\pm$1.12 \\
H2O VMR & 1.39$\pm$1.07 & 1.16$\pm$1.04 & 0.882$\pm$0.947 \\
CO2 VMR & 0.94$\pm$1.18& 0.876$\pm$1.16 & 0.76$\pm$1.09 \\
CH4 VMR & 0.217$\pm$0.634 & 0.169$\pm$0.732  & 0.125$\pm$0.656\\
10-bar Radius & 15455$\pm$52 km & 15320$\pm$58 km & 15190$\pm$59 km\\
\hline
\end{tabular}
\caption{The parameter values in the best-fit models for three
  different atmospheric temperatures, as Table~\ref{cloudtops}. \label{tempresults}}
\end{table}

\begin{table}
\centering
\begin{tabular}[c]{|c|c|c|}
\hline
Variable & Retrieved & Error\\
\hline
Cloud 1 & 2.17 & 2.13\\
Cloud 2 & 0.992 & 0.992 \\
H2O VMR &500 & fixed\\
CO2 VMR &1.00 & 1.00 \\
CH4 VMR &0.923 & 0.896 \\
10-bar Radius & 16634 & 18 km\\
\hline
\end{tabular}
\caption{The parameter values in the retrieved model for a 50\%
  H$_2$O atmosphere, as Table~\ref{cloudtops}. The H$_2$O VMR is set,
  not retrieved, and a full bracketed retrieval was not performed in
  this case, so the error is the retrieval error from a single run.\label{h2o50}}
\end{table}

\begin{figure}
\includegraphics[width=0.5\textwidth]{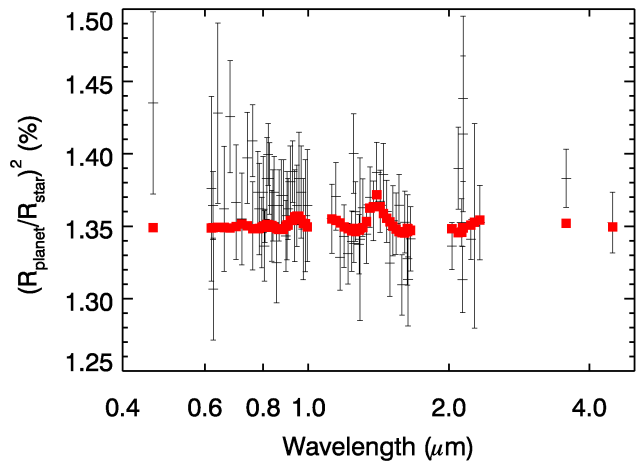}
\caption{Best-fit spectrum for GJ 1214b where the H$_2$O VMR is set
  to 0.5. The data are shown in black and the model is shown in
  red.\label{h2o_dominated}}
\end{figure}

The effect of changing the cloud top altitude on the results can
clearly be seen, and for the most part is straightforward to
understand. Decreasing the cloud top pressure and increasing the
altitude means that gas absorption features are truncated at
higher altitudes, so abundances must increase in order to fit the size
of the observed features, and the radius at 10 bars
is smaller because the higher cloud increases the radius of
atmospheric extinction. The opposite is true when the cloud top
pressure is increased/altitude is decreased. Less intuitively, the
cloud abundances decrease if the cloud top pressure is either lowered
or raised, indicating the complexity of the degeneracy between
different scenarios; for example, the CH$_4$ VMR is very low for
the 10 mbar case, so the features appear to be the same size despite
both a lower cloud top and a lower abundance. It is clear that
degeneracies allow compensation between the model parameters such that
a reasonable fit to the data can be achieved regardless of the
position of the cloud top.

These results indicate that different assumptions about the vertical
distribution of cloud have an effect on retrieval results; the most
significant effect is on the cloud 2 abundance, which varies by a
factor of 5 if the cloud top is moved from 1 mbar to 0.1
mbar. With the current quality of data we cannot expect to achieve better
retrieval precision than this anyway, but when the data are
more constraining this kind of degeneracy will limit our ability to
draw firm conclusions about GJ 1214b. A comparison between
retrieval results and \textit{ab initio} models could allow
differentiation between degenerate atmospheric scenarios on the basis
of physical likelihood. 

\begin{figure*}
\centering
\includegraphics[width=0.8\textwidth]{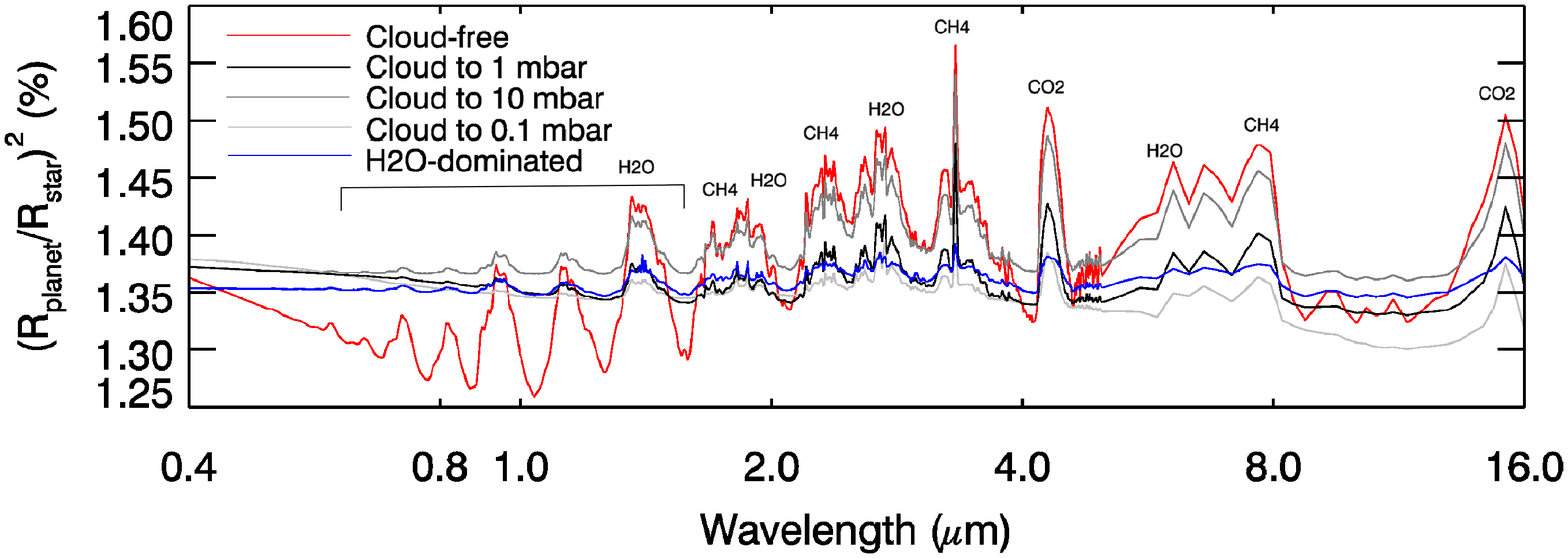}
\caption{Synthetic spectra between 0.4 and 16 $\upmu$m for a range of
  the cases discussed in this paper, with the major gaseous absorption
  bands indicated. It can be seen that
  cloudy/cloud-free/water-rich models are very different at short
  wavelengths, and that the shapes of the CH$_4$ band at 3.3 $\upmu$m
  and the CO$_2$ bands at 4.3 and 16 $\upmu$m will also be very important,
  if these molecules are present in the atmosphere. Spectra for
  different temperatures are not shown because they are very similar
  and therefore still degenerate even with greater spectral coverage,
 so it will be necessary to obtain emission spectra to constrain
temperature structure.\label{models}}
\end{figure*}

\begin{figure*}
\centering
\includegraphics[width=0.8\textwidth]{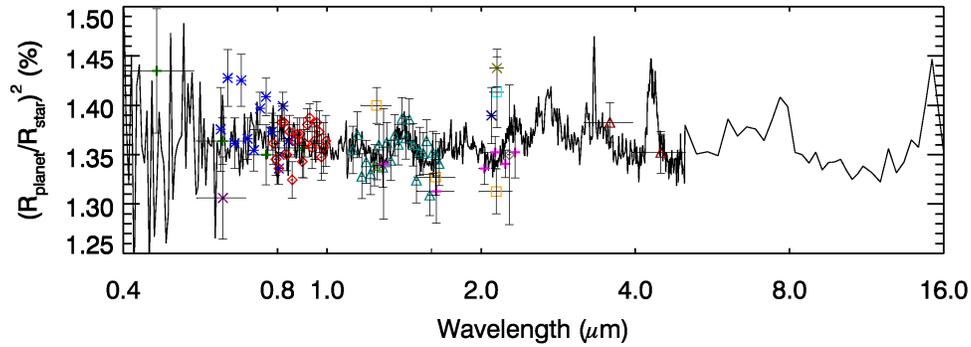}
\caption{Our best-fit model spectrum as it would be seen by EChO, with
  the current data also plotted as in Figure~\ref{data_plot}. The
  noisy synthetic has been generated as in \citet{barstow13},
  assuming photon noise and 30 coadded transits. Whereas the
  faintness of the M dwarf star at short wavelengths means the
  spectrum is noise-dominated here, the coverage in the infrared would
  prove very useful. \label{gj1214becho}}
\end{figure*}

\begin{figure*}
\centering
\includegraphics[width=0.8\textwidth]{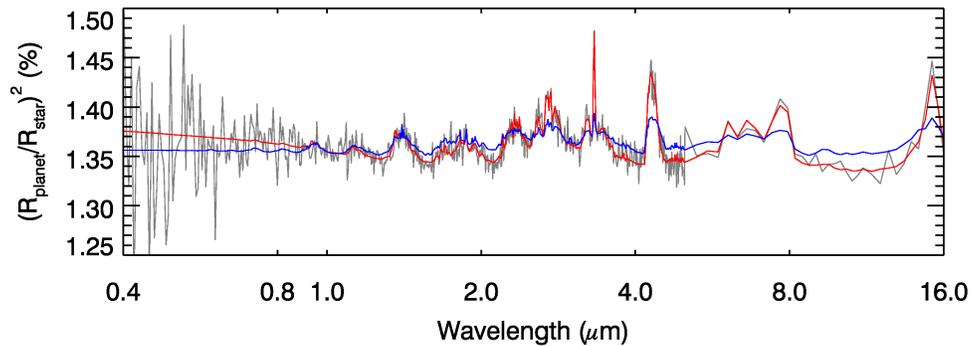}
\caption{Two retrieval fits to the noisy EChO synthetic, with the H$_2$-He model
  (red) and a 50\% H$_2$O model (blue). It can be seen clearly that
  the 50\% H$_2$O model does not produce an adequate fit to a noisy
  EChO synthetic generated with a H$_2$-He model atmosphere. This
  indicates that with EChO we should be able to distinguish between
  the two competing scenarios.\label{echoretrieved}}
\end{figure*}

\subsection{Temperature profile}
\label{tempvary}
Whilst there is insufficient information in the spectrum to retrieve the atmospheric
temperature, the atmospheric scale height is proportional to
temperature so it will have some effect on the spectrum. We test this
by repeating the retrieval with our input temperature profile shifted
by -50 and +50 K. As in section~\ref{cloudalt}, we
present the best-fit parameter values from each case in
Table~\ref{tempresults}. The lower temperature profile
tested here corresponds to an equilibrium temperature of 480 K and a
Bond albedo of 0.5 in our simple temperature profile model, which may
be a likely scenario if the
planet's albedo is dominated by scattering from a high, reflective
cloud layer. 

A decrease in temperature decreases the atmospheric scale height,
which also decreases the radius of the atmosphere above the 10 bar
pressure level. The 10-bar radius and cloud opacities therefore
increase to counteract this effect. A decreased scale height also
makes gaseous absorption features appear to be flatter, so the gas
VMRs are increased to compensate for this. The opposite is true
when the temperature is increased. A very similar effect to that for a temperature decrease would be observed if the mean molecular weight
of the atmosphere was increased by the addition of a heavier spectrally-inactive gas
such as N$_2$.

As in Section~\ref{cloudalt}, changing the model temperature
profile has a non-negligible effect on the retrieved
parameters, so the radius retrieval alone does not fully compensate for
the effect of temperature on the atmosphere scale height. Again, with improved data quality and better constraints
this inherent degeneracy will
become more important. It will therefore be essential to
observe a secondary transit of GJ 1214b in the future, as this could
provide further information about the temperature structure of the
atmosphere. 

\subsection{H$_2$O-dominated atmospheres}
\label{h2odominated}

We have not considered in detail the possibility of H$_2$O-dominated
atmospheric scenarios, since it has already been shown
\citep{bean11,berta12} that a
reasonable $\chi^2$ can be achieved with a high molecular weight
atmosphere. For completeness, we perform a retrieval with the H$_2$O
VMR set to 0.5, varying the cloud abundances, CO$_2$/CH$_4$ volume
mixing ratios and the radius. A fit can be achieved with a reduced
$\chi^2$ of 0.92, for the parameter values in Table~\ref{h2o50}.

The best-fit model for the 50\% H$_2$O scenario is shown in
Figure~\ref{h2o_dominated}. It can be seen that the features do not
differ greatly from the best-fit H$_2$-He models, except the slight
increase in radius towards the blue end of the spectrum is no longer
seen for a high molecular weight atmosphere. This demonstrates that
our method can also produce a good fit to the spectrum for an
H$_2$O-dominated atmosphere, and so despite showing that a cloudy
mini-Neptune atmosphere is a strong possibility we cannot rule out the
water-world scenario for GJ 1214b.

\subsection{Future measurements}

The data currently available are sparse and have low
signal-to-noise. Future missions such as the James Webb Space
Telescope and the Exoplanet Characterisation Observatory (EChO,
\citealt{tinetti11}) will enable the whole
near-infrared spectrum to be covered simultaneously at high precision
and, in the case of EChO, the full spectrum between 0.55 and 16
$\upmu$m (Figure~\ref{gj1214becho}). In
addition, ground-based techniques for transit spectroscopy, particularly those employing
multi-object spectroscopy (e.g. \citealt{bean10,gibson13}), are constantly
improving. High-dispersion spectroscopy techniques such as that
pioneered by \citet{snellen10} can provide unambiguous detection of
molecular absorbers in exoplanet atmospheres, which will also help to
break degeneracies. With higher precision and better coverage, future space- and
ground-based observations should enable us to finally
distinguish between competing scenarios for this planet, and break
some of the degeneracies explored in this paper. We show a range of
synthetic spectra for the spectral range and resolution probed by EChO
to demonstrate this (Figure~\ref{models}); clear differences between
the scenarios can be seen in the visible, and in CH$_4$ and CO$_2$
absorption bands in the infrared. However, spectra with different
atmospheric temperature structures are still degenerate even with
increased coverage and spectral resolving power, so it would be
necessary to observe a secondary transit to fully constrain the
properties of the atmosphere.

We investigate whether there is sufficient information to constrain
the atmosphere in an EChO
spectrum by
performing the same analysis as that described in
\citet{barstow13}. We take the synthetic spectrum calculated from the
best-fit model atmosphere and add the expected level of Gaussian
random noise for an
EChO observation of GJ 1214b, as shown in Figure~\ref{gj1214becho}. We then
feed the noisy spectrum back into NEMESIS to perform a retrieval, as
though it was an observed spectrum, and then compare the
retrieved parameters with the original model parameters (Table~\ref{echoretrieval}). We find that
NEMESIS could retrieve the H$_2$O and CH$_4$ VMRs from a noisy EChO spectrum
to within 10\% of the input value; the cloud 1 number density is
retrieved to within 15\%, and all these parameters are retrieved to within 1$\sigma$, given correct
estimates for the cloud top height and temperature profile. The cloud
2 number density and CO$_2$ VMR were retrieved correctly to within 2$\sigma$. 

\begin{table}
\centering
\begin{tabular}[c]{|c|c|c|c|}
\hline
Variable & Input & Retrieved & Error \\
\hline
Cloud 1 & 1.55 & 1.31 & 0.24\\
Cloud 2 & 0.783  & 1.10 & 0.25\\
H2O VMR & 1.16 & 1.28 & 0.21\\
CO2 VMR & 0.876 & 1.14 & 0.17\\
CH4 VMR & 0.169  & 0.177 & 0.022\\
10-bar Radius & 15320 km & 15316 km & 8 km\\
\hline
\end{tabular}
\caption{The retrieved parameter values for the synthetic noisy EChO
  spectrum compared with the known input values. The \textit{a
    priori} values were all 1.0, with the exception of the radius
  for which it was 15455 km.\label{echoretrieval}}
\end{table}

\begin{table}
\centering
\begin{tabular}[c]{|c|c|c|}
\hline
Wavelength ($\upmu$m)& Reduced $\chi^2$ (H$_2$-He) & Reduced $\chi^2$ (H$_2$O) \\
\hline
0.55---0.95 & 1.08 & 1.14\\
3---5  & 0.91  & 2.96 \\
5---11& 1.53 & 8.49\\
5---16 & 1.29 & 7.28\\
Full & 0.90  & 2.10\\

\hline
\end{tabular}
\caption{Reduced $\chi^2$ for different spectral ranges, comparing
  the fit of the H$_2$-He model and the 50\% H$_2$O model with the
  noisy synthetic EChO spectrum.\label{chi2}}
\end{table}

We perform the same retrieval test with the noisy EChO synthetic when
the H$_2$O VMR is forced to be 50\%. The retrieved spectrum does not
provide as good a fit to the noisy synthetic as the retrieved H$_2$-He
atmosphere spectrum, giving a reduced $\chi^2$ of 2.1 instead of
0.9. The two retrieved spectra are shown overplotted on the noisy
synthetic in Figure~\ref{echoretrieved}, and it can seen that the
regions in which they differ most are the visible region, the infrared
longwards of 5 $\upmu$m, and the CH$_4$ and CO$_2$ bands at 3.3 and
4.3 $\upmu$m respectively. The EChO spectrum for GJ 1214b is likely to
be too noisy in the visible to distinguish between the models, as
demonstrated by comparable reduced $\chi^2$ over the range from
0.55---0.95 $\upmu$m (Table~\ref{chi2}), but it is clear that the
CH$_4$/CO$_2$ bands and especially the mid-infrared region are
very useful for distinguishing between the two scenarios with
EChO. With data of the coverage and quality we expect from a space
telescope such as EChO, we
should be able to distinguish between the cloudy mini-Neptune and
water-world scenarios for GJ 1214b; however, our abilty to provide more
detailed constraints on the atmosphere is limited by the degeneracies
discussed in Sections~\ref{cloudalt}
and~\ref{tempvary}. 

\section{Conclusions}
\label{conclusions}

We have used the NEMESIS radiative transfer and retrieval tool to
explore the degeneracy of the retrieval problem for GJ 1214b.  We find that the spectroscopic data are compatible with an
H$_2$-He dominated, cloudy atmosphere. A range of models with 0.1
$\upmu$m tholin haze particles, 1 $\upmu$m tholin cloud particles and trace amounts of
H$_2$O, CO$_2$ and CH$_4$ produce synthetic spectra that provide a
good fit to the data; however, the number of models with a good fit
allow for several orders of magnitude of variation
in the abundances of these, so it is difficult to place meaningful
constraints. We also cannot rule out the possibility of an
H$_2$O-dominated atmosphere with a small scale height, as this results
in a synthetic spectrum with an equally good fit to the data. In addition, for a cloudy H$_2$-He atmosphere the cloud top pressure and
temperature profile specified in the model atmosphere significantly effect the
retrieved cloud and gas abundances, indicating the presence of further
model degeneracy. A disc-integrated
emission spectrum from secondary transit will help to constrain the
temperature profile, and will be necessary to break these degeneracies.

Future observations will be crucial for finally determining the nature
of GJ 1214b. Improvements in the precision of
ground-based transit spectra are hoped to provide more conclusive
answers, and in the longer term we look to space-based missions such
as EChO, which should be able to distinguish between H$_2$-He- and
H$_2$O-dominated atmospheres. We have demonstrated that NEMESIS is a
valuable tool, and our
exploration of the degeneracies in this retrieval problem will
enable us to find the best approach for the interpretation of future data.

\section*{Acknowledgements}
JKB acknowledges the support of the John Fell Oxford University Press
(OUP) Research Fund for this research and LNF is supported by a Royal
Society Research Fellowship. We thank the anonymous reviewer for their
comments on this paper. 

\bibliographystyle{mn2e}
\bibliography{bibliography}

\label{lastpage}
\end{document}